# Crystal growth and magnetic properties of $Nd_{1-x}Dy_xFe_3(BO_3)_4$

I.A. Gudim, E.V. Eremin, V.L. Temerov

*Kirensky Institute of Physics SB RAS, Academgorodok, Krasnoyarsk, Russian Federation*

e-mail: bezm@iph.krasn.ru

Data about stability areas of $Nd_{1-x}Dy_xFe_3(BO_3)_4$ trigonal phases in fluxes on the bismuth trimolibdate basis and feature of labile phase dynamics conditions is noted. Group growth conditions of these phases single crystals on seeds and at spontaneous nucleation in a thin layer of flux on crystal carrier are described.

Temperature and field dependences of magnetization and a specific heat for crystals with x=0,1 and x=0,25 are analyzed. The conclusion, that unlike easy plane $NdFe_3(BO_3)_4$ antiferromagnetic and easy axis $DyFe_3(BO_3)_4$ ones, spontaneous spin-flop transitions are inherent in these single crystals is resulted and it is connected with mutual influence of rare-earth subsystems through the common Fe-subsystem with strong exchange interaction. The anisotropies competition exert essential influence on induced by field H ‖$c_3$ magnetic phase transitions too.

1. Introduction.

With detection of sufficiently strong electric polarizations, induced by a magnetic field, in trigonal $GdFe_3(BO_3)_4$ [1] and $NdFe_3(BO_3)_4$ [2] antiferromagnetics, interest to magnetic and magnetoelectric properties complex research of others isostructural rare-earth ferroborates $RFe_3(BO_3)_4$ has considerably increased. For clarification of magnetoelectric polarization microscopic picture in such structures it is represented important to experiment on monocrystals with a double rare-earth subsystem, including solid solutions $NdFe_3(BO_3)_4$ and $DyFe_3(BO_3)_4$ [3], representing a combination of two antiferromagnetics with different magnetocrystallic anisotropy - "an easy plane" and "an easy axis". The main element of rare-earth ferroborates crystal structure is the spiral chains of $FeO_6$ octahedrons was contiguous on an edge and oriented along the *c* axis [4]. Antiferromagnetic interaction between $Fe^{3+}$ ions along a chain is stronger, than ferromagnetic ordering between chains. Rare-earth ferroborates magnetic properties are defined both an iron subsystem of ions $Fe^{3+}$, and a rare-earth subsystem of ions $R^{3+}$. The iron subsystem in $RFe_3(BO_3)_4$ is ordered at temperatures of Neele $T_N \approx$ 30-40K and can be considered as set of two antiferromagnetic sublattices. The rare-earth subsystem remains paramagnetic up to temperatures ≈ 1K, but ones magnetic biased because of f – d interactions and as can be considered as two sublattices. It is known that the rare earths are responsible for magnetic anisotropy and orientation of the magnetic moments in these compounds. So in $NdFe_3(BO_3)_4$ ferroborate [2,5] the easy-axis magnetic structure is realized below temperature of Neele $T_N$ =32K, whereas in $DyFe_3(BO_3)_4$ ferroborate [6] the magnetic structure is realized easy-plane below $T_N$ = 38K. There are no spontaneous spin-flop transitions in these compounds. The transitions induced by a magnetic field along an *c* axis for $DyFe_3(BO_3)_4$ [6] and along *a* and *b* axes for $NdFe_3(BO_3)_4$ [7] however take place. For compounds contain simultaneously Dy and Nd ions the magnetic structure will be defined by a competition of «an easy axis» - «an easy plane» type anisotropies.

In this paper the questions connected with control crystal nucleation of trigonal phases $Nd_{1-x}Dy_xFe_3(BO_3)_4$ (0 <x <1) in $Bi_2Mo_3O_{12}$ – $B_2O_3$ – $Nd_2O_3$ - $Dy_2O_3$ – $Fe_2O_3$ flux system are considered. The special attention is given to metastability and phase dynamics of labile states on border with accompanying phases. The technique of single crystals $Nd_{1-x}Dy_xFe_3(BO_3)_4$ group growth on seeds is offered. Temperature and field dependences of magnetization and specific heat of single crystals $Nd_{1-x}Dy_xFe_3(BO_3)_4$ with x=0,1 and x=0,25 are discussed.

2. Phase formation.

The investigated flux system is more convenient to write in the quasibinary form n%mass. [$Bi_2Mo_3O_{12}$ + p$B_2O_3$ + q$Nd_2O_3$ + r$Dy_2O_3$] + n%mass.$Nd_{1-x}Dy_xFe_3(BO_3)_4$ where n – is a crystal-forming oxides concentration, corresponding to $Nd_{1-x}Dy_xFe_3(BO_3)_4$ stoichiometry, and p, q, r parameters indicate the contain of crystal-forming oxides over stoichiometry. Fluxes with various p, q, r, and n values and weight of 300g are prepared at T=1000-1100$^0$C by consecutive oxides sequential melting [$Bi_2O_3$ (reagent grade), $MoO_3$ (analytical grade)], $B_2O_3$ (special-purity grade), [$Fe_2O_3$ (special-purity grade), $Nd_2O_3$ (NdO-E), $Dy_2O_3$ (DyO-L)] (domestic reagents) in platinum cylindrical cruisible with a 50mm diameter and 60mm height. Then the crucible with the prepared flux was located into the furnace with the temperature field which vertical component at T=1000$^0$C decreases with a gradient 2-3$^0$C/cm at removal from a bottom cruisible. After flux homogenization at T=1000$^0$C within 24 hours the temperature was lowered to the expected saturation temperature ($T_{sat}$) and a platinum crystal-carrier in the form of a core in diameter 4mm was dipped into the flux. In 1-2 hours crystal-carrier it was withdrawn and the nucleation on it was estimated. Further this probes were proceeded with a 10-20$^0$C temperature fall step, down to T≈850$^0$C without flux overheat.

It is established that hematite (α-Fe2O3) is solo crystallizing phase at p = q = r =0 and n=10÷25 in the range from its $T_{sat}$ to 850$^0$C. With increase of $B_2O_3$ and $(Nd,Dy)_2O_3$ content, accordingly to p=2 and (q+r) =0,4, hematite remains a high-temperature phase, and $Fe_3BO_6$ crystallite appear in the lower part of a temperature interval. Trigonal $(Nd,Dy)Fe_3(BO_3)_4$ becomes a high-temperature phase with a wide stability range only at p=2,5-3 and (q+r) =0,5-0,6. α-$Fe_2O_3$ and $Fe_3BO_6$ phases keep the tendency to formation, but they appear near to the lower border of $Nd_{1-x}Dy_xFe_3(BO_3)_4$ metastability zone, and in some time after formation they start to be dissolved ("nonequilibrium effect"). Nonequilibrium effect and time of its relaxation was estimated, observing behavior of the equilibrium and labile phases simultaneously dipped in the flux. At temperature on 12$^0$C below $Nd_{1-x}Dy_xFe_3(BO_3)_4$ formation, the appreciable dissolution of α-$Fe_2O_3$ and $Fe_3BO_6$ was observed in 10-15 hours, and with overcooling increase to 20$^0$C this time exceeded 1 day. The overcooling threshold value, since which any or both simultaneously labile phases were nucleated, essentially depended on parameters p, q, and r. That additionally narrowed a range of a possible choice of values p, q, r, and demanded acceptance of measures to a non-admission of their appreciable deviation during flux preparation, particularly, because of uncontrollable composition change at the evaporation.

Main crystallization parameters for the fluxes with p=3 and (q+r) =0,6 are resulted in Table1.

Table 1

| Flux composition in the quasibinary form n%масс.[$Bi_2Mo_3O_{12}$ + p$B_2O_3$ + q$Nd_2O_3$ + r$Dy_2O_3$] + n%масс.$Nd_{1-x}Dy_xFe_3(BO_3)_4$ | | | | | | | | |
|---|---|---|---|---|---|---|---|---|
| Crystal | p | q | r | n | $T_{sat}$, $^0$C | $dT_{sat}/dn$, $^0$C/%mass. | $\Delta T_{met}$, $^0$C | Accompaning phases |
| $NdFe_3(BO_3)_4$ | 3 | 0,6 | 0 | 25 | 962 | 2,5 | ≈12 | α-$Fe_2O_3$, $Fe_3BO_6$ |
| $Nd_{0,95}Dy_{0,05}Fe_3(BO_3)_4$ | 3 | 0,57 | 0,03 | 25 | 965 | 4,0 | | |
| $Nd_{0,9}Dy_{0,1}Fe_3(BO_3)_4$ | 3 | 0,54 | 0,06 | 25 | 985 | 4,8 | | |
| $Nd_{0,85}Dy_{0,15}Fe_3(BO_3)_4$ | 3 | 0,51 | 0,09 | 25 | 980 | 3,6 | | |
| $Nd_{0,75}Dy_{0,25}Fe_3(BO_3)_4$ | 3 | 0,45 | 0,15 | 25 | 968 | 3,0 | | |
| $DyFe_3(BO_3)_4$ [8] | 3 | 0 | 0,5 | 30 | 980 | 4,5 | | |

At crystal-forming oxides concentration of n=25% the saturation temperature is smaller 1000$^0$C, that uncontrollable crystallization condition change are excluded thereby at long-term works with opened cruisible. The specified width of a metastable zone $\Delta T_{met}$≈12$^0$C corresponds to 20 hour endurance of a flux in the overcooled state. With increase of endurance time it did not vary practically and consequently it is possible to consider it long-term. At temperatures lower than 870$^0$C the crystallization is ineffective because of increasing viscosity of the flux and permissible overcooling (narrowing $\Delta T_{met}$).

3. Single crystals growth. The "nonequilibrium effect" account.

Initial crystals growth process in spontaneous formation regime can be conventionally divided on two stages. On the first stage, at flux overcooling to $\Delta T > T_{sat} - \Delta T_{met}$, the probability of formation and growth rate of labile phases are above, than $Nd_{1-x}Dy_xFe_3(BO_3)_4$. It causes the reduction of $Nd_{1-x}Dy_xFe_3(BO_3)_4$ crystallite nucleation quantity, to their growth rate slowed and possible accretion with accompanying phase crystallite. At the second stage the dissolution of labile phases begins and $Nd_{1-x}Dy_xFe_3(BO_3)_4$ growth rate increases also that faster, than grater the initial overcooling has been set.

Since the quantity of growing crystals in the regime with spontaneous nucleation is not known, it is impossible to set optimum subsequent temperature lowered rate. The crystals growth process in regime with use of seeds is most controllable. In this case initial overcooling and subsequent temperature lowered rate should be set so that critical overcooling $\Delta T_{init} < \Delta T_{crit} = 10-15^oC$ was not reached. Infringement of this condition can result to competing phase formation, both on crystal-carrier, and at a cruisible bottom and walls. In both cases crystallization control is lost. On the other hand realization of a regime with labile phase preliminary spontaneous crystallization is not excluded. Thus seeds growth rates will be more, while their growth is feeded with a dissolved labile phase.

### 3.1. Seeds production.

At $T=1000^0C$ the crystal-carrier was dipped into flux, and it is recessively rotated at 30 rpm. The furnace temperature was lowered to $T=T_{sat}-12^0C$. In 2 hours crystal-carrier it was withdrawn from the furnace. Thus crystallines of a cooled flux enveloping crystal-carrier were formed on it. Than the crystal-carrier was again dipped into the flux (without overheat at the same temperature $T= T_{sat}-12^0C$) and rotated at 30 rpm with every minute reveres. In the next 24 hours 10-30 crystals with 0,5 - 2mm size grew on it. They had high quality, and after withdrawal were used as seed material.

### 3.2. Crystal growth on seeds.

The crystal-carrier with four qualitative seeds ~1 mm in size was suspended above flux at $T=1000^0C$. After temperature decreased to $T=T_{sat} + 7^0C$ crystal-carrier with seeds was dipped into the flux on 15 - 20 mm depth and reverse rotation with a speed of 30 rpm and a 1 minute period was switched on. After 15 minutes the temperature decreased to $T=T_{sat} - 5^0C$. Furthermore, temperature gradually decreased at $1^oC$/day rate. In 9-13 days the growth process was finished. Crystal-carrier was lifted above flux and cooled to a room temperature with the furnace supply switched off. As a result crystals about the 6 – 10 mm sizes sufficient for carrying out of physical researches are grown up. Spontaneous formation and other phases were absent. Result is present in fig. 1.

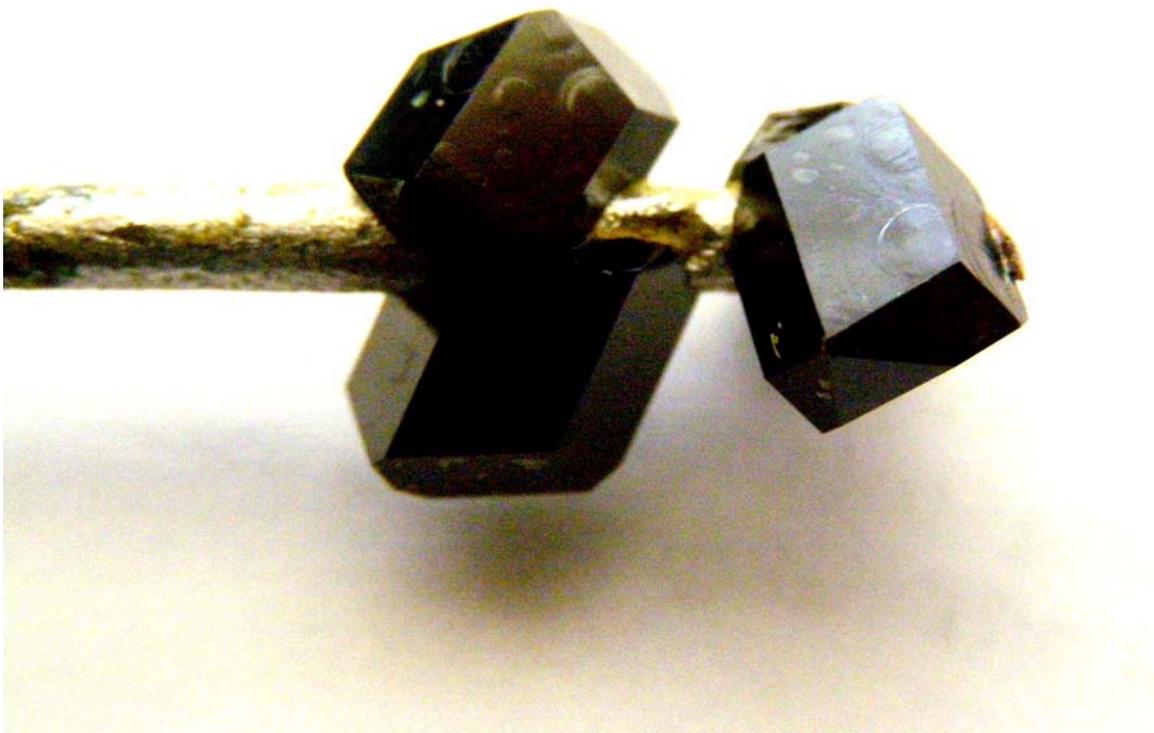

Fig. 1. $Nd_{0.75}Dy_{0.25}Fe_3(BO_3)_4$ single crystal grown from bismute molibdate based flux.

    4. Magnetic and calorimetric properties.

All temperature and calorimetric measurements have been spent on Physical Properties Measurement System (Quantum Design) in a temperature interval 2-300K and magnetic fields to 9T. On fig. 2 the temperature dependences of magnetization of M (T) for $Nd_{0.9}Dy_{0.1}Fe_3(BO_3)_4$ and $Nd_{0.75}Dy_{0.25}Fe_3(BO_3)_4$ compounds carry out for magnetic field $H_{DC}=1$кOe with orientation along and perpendicularly to third order axes are resulted. At temperature decrease at T ≈ 31K in both compounds three-dimensional antiferromagnetic ordering takes place. For compound

$Nd_{0.9}Dy_{0.1}Fe_3(BO_3)_4$ magnetization along the *c* axis fails with temperature increasing continues,

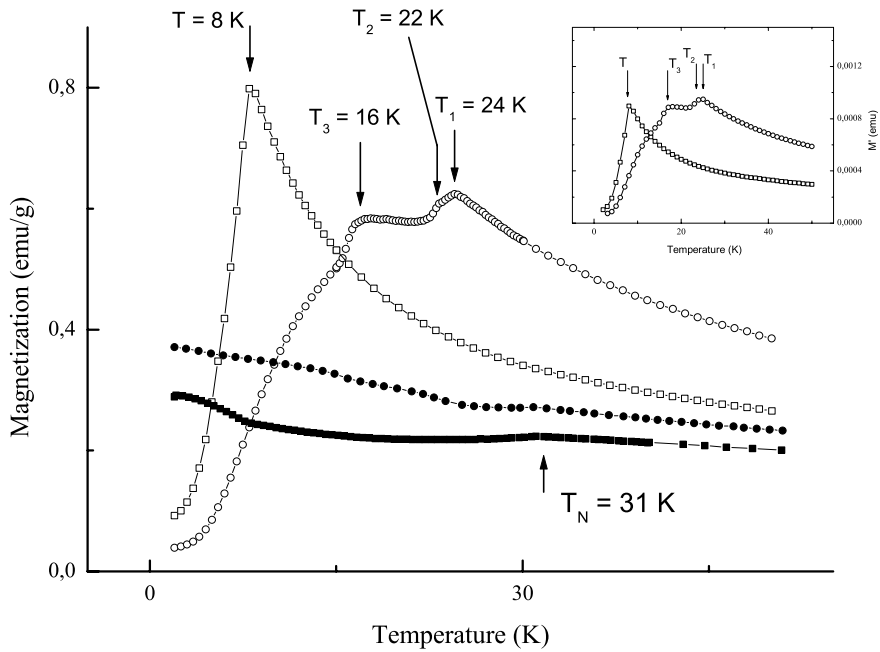

Fig. 2. Temperature dependences of magnetization for $Nd_{1-x}Dy_xFe_3(BO_3)_4$ (x=0,1 and x=0,25) at H=1kOe. An insert - temperature trend of the real part M′(T). ○,● – x=0,25; □,■ – x=0,1.

that is characteristic for easy-plane states of magnetic structure. At the further reduction of temperature from T ≈ 8K magnetization begins to decrease sharply that, apparently, corresponds to spontaneous transition in uniaxial state. A temperature trend of the real part M′(T) (an insert on fig. 2) is evident to spin-flop transition spontaneous character. Field dependences of magnetization M (H) at various temperatures for this structure are presented on fig. 3.

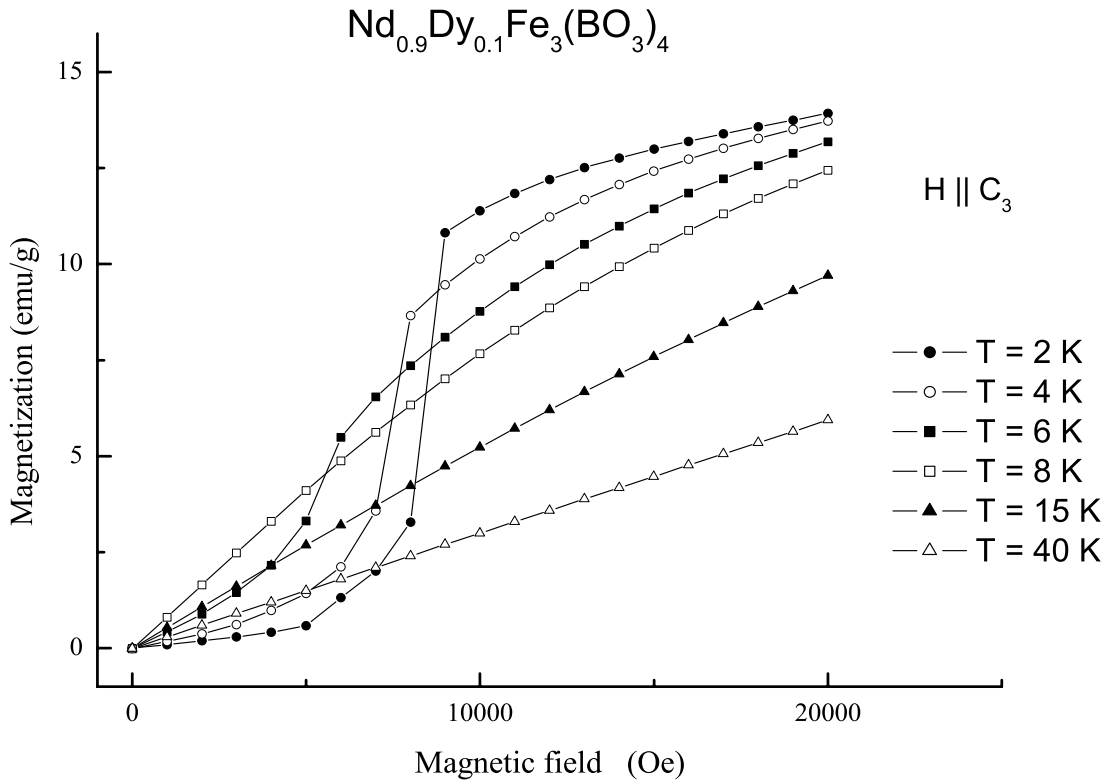

Fig. 3. Field dependence of magnetization for

It is visible that at temperatures of lower than 8K the magnetizing field $H_{DC}//c$ increase conducts to jump-increase of magnetization, that corresponds to spin-flop transition from uniaxial state in the easy-plane state. It is interesting that the spin-flop field increases with temperature lowered, whereas, for example, for $DyFe_3(BO_3)$ [6] it decreases with temperature pull.

For $Nd_{0,75}Dy_{0,25}Fe_3(BO_3)_4$ compound M (T) behavior becomes much more difficult (fig. 2). Reorientation from easy-plane state in uniaxial ones takes place here too. However on magnetization temperature dependence trend along the c axis three anomalies are visible - at $T_1 \approx$ 24K, $T_2 \approx$ 22K and $T_3 \approx$ 16K. Features in temperature dependence of a thermal capacity $C_p$ (fig. 4) are observed at the same temperatures. It is necessary to notice that if at $T_1$ and $T_2$ features on a curve thermal capacity has narrow enough peak, at $T_3$ peak is strongly smearing. This feature at $T_3$ can be caused by Shotky anomaly which was also observed both for $DyFe_3(BO_3)_4$ [6], and for

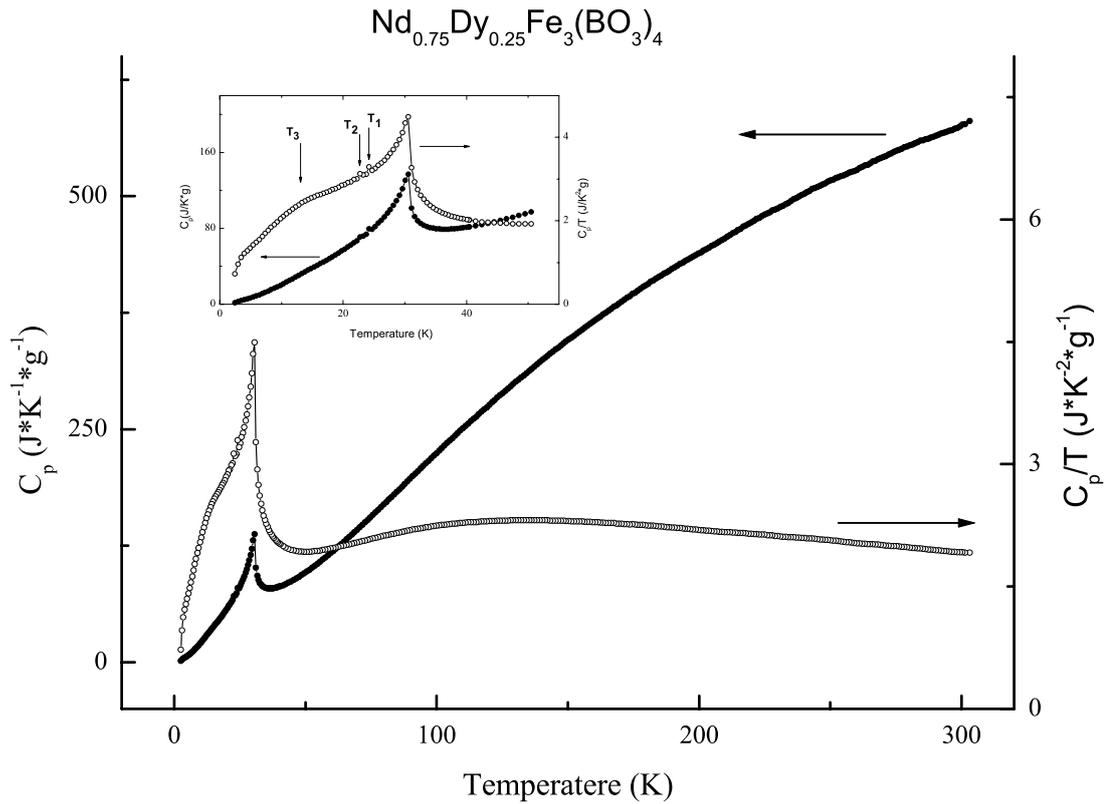

Fig. 4. The temperatures dependence of specific heat for $Nd_{0.75}Dy_{0.25}Fe_3(BO_3)_4$.

$NdFe_3(BO_3)_4$ [5]. On fig. 5 field dependences of magnetization of M (H) measured for $Nd_{0.75}Dy_{0.25}Fe_3(BO_3)_4$ are presented at various temperatures. It is visible from figure that the transitions from uniaxial state in the easy-plane ones induced by a magnetic field are also realizes difficultly. For example, at T ≈ 22K there are three jumps that are well visible from an insert on fig. 5.

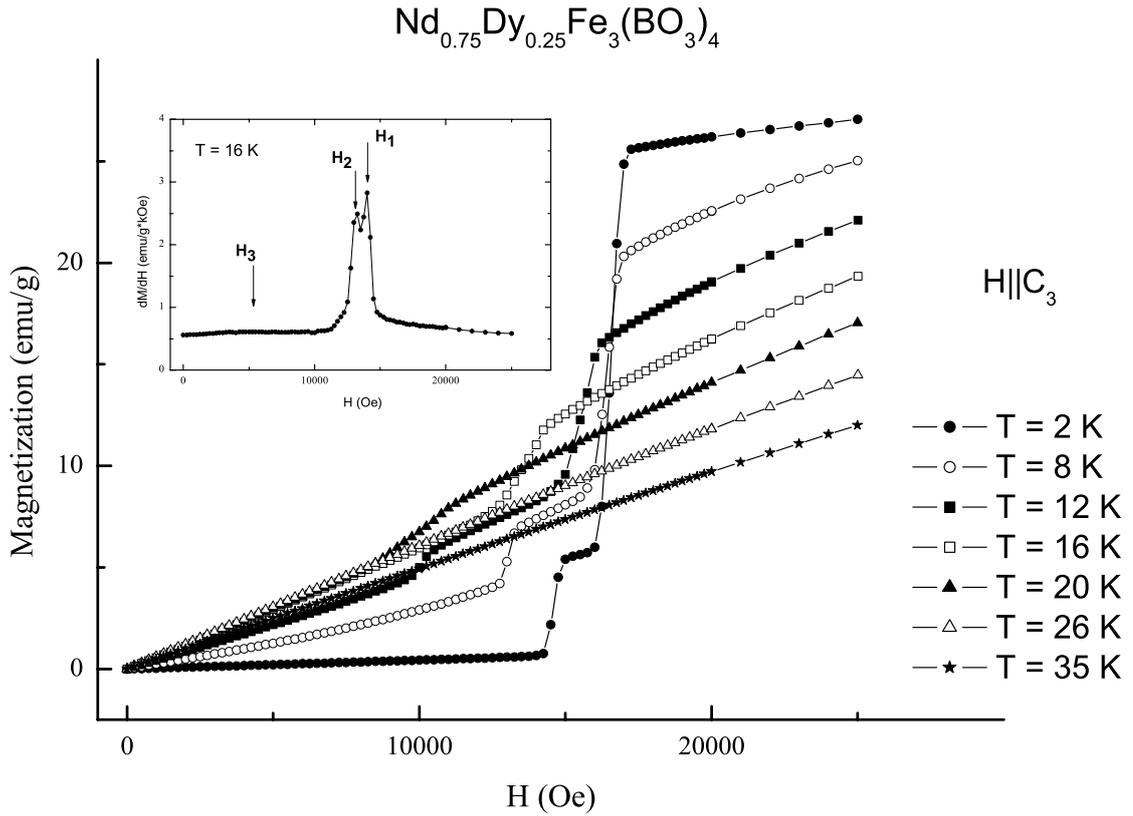

Fig. 5. Field dependence of magnetization for $Nd_{0,75}Dy_{0,25}Fe_3(BO_3)_4$.

Thus, in $Nd_{1-x}Dy_xFe_3(BO_3)_4$ crystals depending on rare-earth subsystem composition it can be realized not only easy-plane (x→0) or easy-axis (x→1) antiferromagnetic state, but at a certain parity of ions $Nd^{3+}$ and $Dy^{3+}$ the situation with occurrence spin reorientation in the $T<T_N$ magnetic ordering area is possible. And even the small Dy content (x=0,1) leads to occurrence of the spin-flop phase transitions spontaneous and induced by a field. With Dy content increase (x=0,25) the temperature of spontaneous spin-flop transition raises. Thus magnetization jumps at T≈16K and T≈24K, apparently, are caused accordingly Nd and Dy. Metamagnetic transition, that is induced by magnetic field H//c, is displaced towards high fields.

5. The conclusion.

In fluxes (100-n)% mass. $(Bi_2Mo_3O_{12} + pB_2O_3 + q(1-x) Nd_2O_3 + qxDy_2O_3)$ + n%mass.$Nd_{1-x}Dy_xFe_3(BO_3)_4$ at any $0 \leq x \leq 1$ in a vicinity $p_0 = 3$, $q_0 = 0,6$, $n_0 = 25$ the p, q, n choice is possible at which $Nd_{1-x}Dy_xFe_3(BO_3)_4$ trigonal phases are high-temperature and crystallize in enough wide temperature intervals. In order of controlled crystallization, the choice of flux composition, initial overcooling and temperature pulling rate should exclude nonequilibrium formation of accompanying phases. Taking into account this requirement techniques of reproduced single crystals growth, applicable for all $Nd_{1-x}Dy_xFe_3(BO_3)_4$ ($0 \leq x \leq 1$) family are realized. The magnetic behavior of these single crystals essentially depends from x. So the received samples with x=0,1 and x=0,25 unlike earlier studied extremes of this family – easy-plane $NdFe_3(BO_3)_4$ antiferromagnetic and easy-axis $DyFe_3(BO_3)_4$ ones – have spontaneous spin-flop transitions. This fact, as well as observed displacement field induced H//c transitions depending on x, are connected by that in mixed crystal $Nd_{1-x}Dy_xFe_3(BO_3)_4 = (1-x)NdFe_3(BO_3)_4 + xDyFe_3(BO_3)_4$ competition of anisotropies easy-plane and easy-axis subsystems is occurred out through the common Fe-subsystem with strong exchange interaction.